\documentclass[%
 reprint,
 superscriptaddress,
 amsmath,amssymb,
 aps,
 floatfix,
]{revtex4-2}

\usepackage{graphicx}
\usepackage{dcolumn}
\usepackage{bm}
\usepackage{subfig}
\usepackage{float}
\usepackage{caption}
\usepackage{siunitx}

\usepackage{xcolor}
\newcommand{\orange}[1]{\textcolor{orange}{#1}} 

\begin{document}

\preprint{APS/123-QED}

\title{Indistinguishable MHz-narrow heralded photon pairs from a whispering gallery resonator}

\author{Sheng-Hsuan Huang}
 \email{sheng-hsuan.huang@mpl.mpg.de}
\affiliation{Department of Physics, Friedrich-Alexander-Universität Erlangen-Nürnberg, Staudtstrasse 7/B2, 91058 Erlangen, Germany} 
\affiliation{Max Planck Institute for the Science of Light, Staudtstrasse 2, 91058 Erlangen, Germany}
\author{Thomas Dirmeier}
\affiliation{Department of Physics, Friedrich-Alexander-Universität Erlangen-Nürnberg, Staudtstrasse 7/B2, 91058 Erlangen, Germany} 
\affiliation{Max Planck Institute for the Science of Light, Staudtstrasse 2, 91058 Erlangen, Germany}
\author{Golnoush Shafiee}
\affiliation{Department of Physics, Friedrich-Alexander-Universität Erlangen-Nürnberg, Staudtstrasse 7/B2, 91058 Erlangen, Germany} 
\affiliation{Max Planck Institute for the Science of Light, Staudtstrasse 2, 91058 Erlangen, Germany}
\author{Kaisa Laiho}
\affiliation{German Aerospace Center (DLR e.V.), Institute of Quantum Technologies, Wihelm-Runge-Str. 10, 89081 Ulm, Germany}
\author{\\Dmitry V. Strekalov}
\affiliation{Max Planck Institute for the Science of Light, Staudtstrasse 2, 91058 Erlangen, Germany}
\author{Andrea Aiello}
\affiliation{Max Planck Institute for the Science of Light, Staudtstrasse 2, 91058 Erlangen, Germany}
\author{Gerd Leuchs}
\affiliation{Department of Physics, Friedrich-Alexander-Universität Erlangen-Nürnberg, Staudtstrasse 7/B2, 91058 Erlangen, Germany} 
\affiliation{Max Planck Institute for the Science of Light, Staudtstrasse 2, 91058 Erlangen, Germany}
\author{Christoph Marquardt}
\affiliation{Department of Physics, Friedrich-Alexander-Universität Erlangen-Nürnberg, Staudtstrasse 7/B2, 91058 Erlangen, Germany}
\affiliation{Max Planck Institute for the Science of Light, Staudtstrasse 2, 91058 Erlangen, Germany}

\date{\today}

\begin{abstract}

Hong-Ou-Mandel interference plays a vital role in many quantum optical applications where indistinguishability of two photons is important. Such photon pairs are commonly generated as the signal and idler in the frequency and polarization-degenerate spontaneous parametric down conversion~(SPDC). To scale this approach to a larger number of photons we demonstrate how two independent signal photons radiated into different spatial modes can be rendered conditionally indistinguishable by a heralding measurement performed on their respective idlers. We use the SPDC in a whispering gallery resonator, which is already proven to be versatile sources of quantum states. Its extreme conversion efficiency allowed us to perform our measurements with only \qty{50}{nW} of in-coupled pump power in each propagation direction. The Hong-Ou-Mandel interference of two counter-propagating signal photons manifested itself in the four-fold coincidence rate, where the two idler photons detection heralds a pair of signal photons with a desired temporal overlap. We achieved the Hong-Ou-Mandel dip contrast of \(74\pm 5\%\). Importantly, the optical bandwidth of all involved photons is of the order of a MHz and is continuously tunable. This, on the one hand, makes it possible to achieve the necessary temporal measurements resolution with standard electronics, and on the other hand, creates a quantum states source compatible with other candidates for qubit implementation, such as optical transitions in solid-state or vaporous systems. We also discuss the possibility of generating photon pairs with similar temporal modes from two different whispering gallery resonators.

\end{abstract}

\maketitle


\section{Introduction}
The two-photon interference, named after Hong, Ou and Mandel~(HOM)~\cite{hong1987measurement}, has attracted a great deal of attention since proposed in 1987. This intriguing effect, which does not have an analogue in classical physics, plays a key role in many quantum applications, including boson sampling~\cite{brod2019photonic}, quantum communication~\cite{hu2023progress, lo2012measurement}, and linear optical quantum computing~\cite{kok2007linear}. In order to achieve the desired results, all these applications use beamsplitter networks and quantum interference of several optical modes. This in turn requires the availability of indistinguishable photon pairs that can efficiently be interfered with each other. In addition to the ability to generate indistinguishable photons, the power consumption of the experiment should also be taken into account, as this may be a limiting factor for scalability.

In order to realize HOM interference between heralded states by four-fold coincidence discrimination, it is essential to generate two pairs of photons, which are indistinguishable in spatial, polarization, and temporal~(frequency) modes. Several groups have demonstrated HOM interference via four-fold coincidence counting in previous experiments, including spontaneous parametric downconversion~(SPDC) 
in \(\chi^2\) materials~\cite{halder2007entangling, bruno2014pulsed} and four wave mixing~(FWM) in on-chip ring resonators~\cite{faruque2018chip, alexander2024manufacturable}, optical fibers~\cite{soller2011high, patel2014independent}, and atomic systems~\cite{qian2016temporal, jeong2017quantum}. On the one hand, SPDC is nowadays a commonly used method for generating quantum states because of its simplicity of operation and op-chip integration, which makes it favored in many experiments. However, the bandwidth of the generated SPDC photons is usually a few hundred GHz, which corresponds to temporal shapes having a width of a few femtoseconds. Considering the jitters of typical single photon detectors, which are usually larger that at least ten picoseconds, the temporal purity of the photons cannot be preserved in an experimental realisation and the visibility of the HOM interference will decrease. On the other hand, the bandwidth of the generated photons from four wave mixing in atomic systems can reach a value below \qty{100}{MHz} in certain  experimental arrangements such that the temporal purity can be preserved. However, the tunability of FWM systems is limited to comparatively narrow frequency window around the pump frequency and is strongly dependent on the atomic vapour used in the particular experiment. In addition, all previous works required lasers with at least several milliwatts of pump power, which may cause problems when scaling up the system. Therefore, photon sources with narrow bandwidth and low pump power are needed for realizing large scale quantum optical applications. 

\begin{figure*}[htbp]
    \includegraphics[width = 0.8\textwidth]{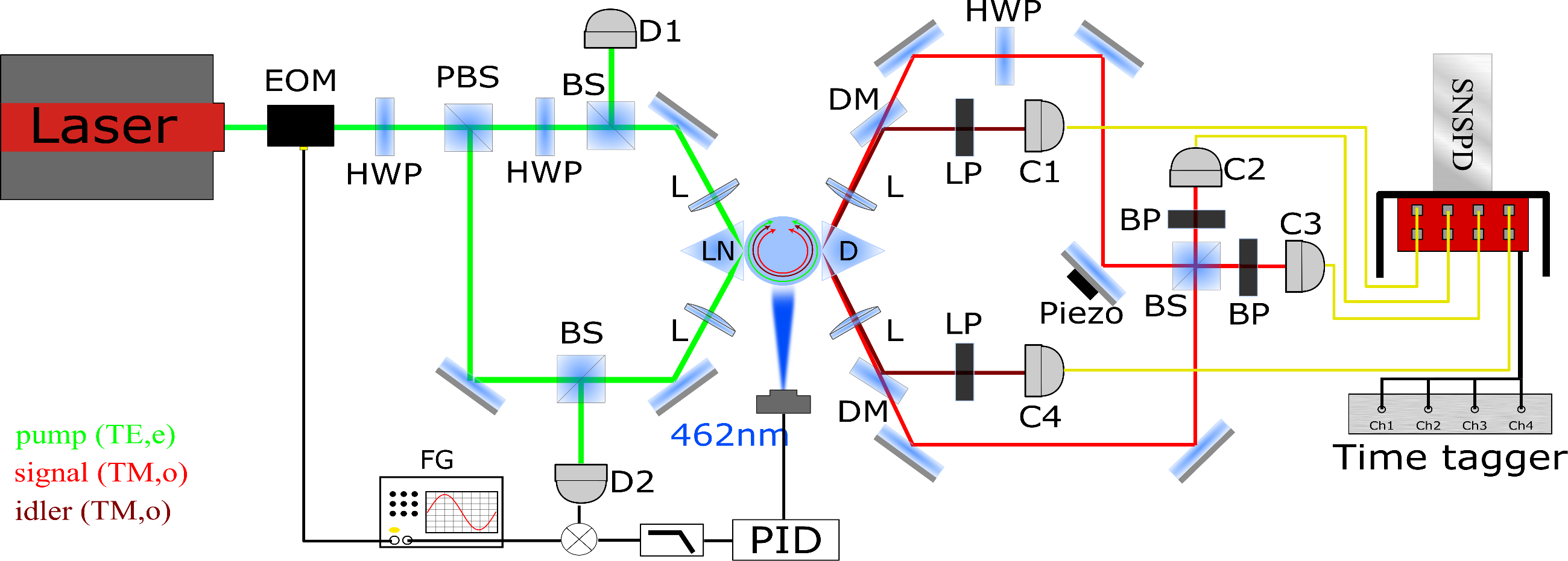}
    \caption{\label{fig:experimental setup}Sketch of the experimental setup. EOM: electro-optic modulator, FG: function generator, HWP: half-wave plate, PBS: polarizing beamsplitter, BS: non-polarizing beamsplitter, PID: proportional–integral–derivative controller, LN: x-cut LiNbO\(_{3}\) prism, D: diamond prism, DM: dichroic mirror, C1-C4: fiber couplers, LP: long pass filter, L:lens, D1-D2: photodetectors, BP:bandpass filter, SNSPD: superconducting nanowire single-photon detector. TM: transverse magnetic modes, TE: transverse electric field modes, o: ordinary, e: extraordinary.}
\end{figure*}

The SPDC in Whispering Gallery Mode Resonators~(WGMRs) has been demonstrated to be a promising source of photonic quantum states, including the generation of heralded single photon states~\cite{fortsch2013versatile}, squeezed states~\cite{otterpohl2019squeezed}, and polarization-entangled photon pairs~\cite{huang2024polarization}. Unlike SPDC in bulk crystals, the bandwidth of the SPDC photons generated from a WGMR is often smaller than \qty{100}{MHz}, providing a sufficient temporal bandwidth that a commercial single photon detector can resolve. The light is trapped in WGMRs due to total internal reflection, so the WGMRs work for the whole transparency region of the material and no special anti-reflection coating is needed. Due to the triply resonant system, the SPDC process is very efficient in WGMRs. The required pump powers are usually smaller than a \(\mu\)W. These features make the WGMRs attractive for large scale quantum information processing as potential sources of quantum states of light. 

In this work, we demonstrate for the first time, HOM interference between heralded states from a WGMR by counting four-fold coincidences. The key requirement for observing the HOM effect is to generate two photons, which are indistinguishable in spatial, polarization and temporal modes. In our case, two photon pairs are generated by coupling the pump beam along the clockwise (CW) and counterclockwise (CCW) directions into the same whispering gallery mode. Since these two counterpropagating beams share the same medium and the same resonant mode, the phase matching conditions are the same for both beams, which makes the two signal photons and the two idler photons generated from the CW and CCW beams identical and traveling in opposite directions. In this experiment, the highest measured four-fold coincidence dip visibility reaches a value of \(74 \pm 5\%\). The in-coupled pump power we used here is only \qty{50}{nW} for each propagation direction. We also show that the HOM dip visibility change when the peak value of the Glauber second-order correlation function, \(g_{\mathrm{si}}^{(2)}(0)\), between signal~(s) and idler~(i) varies, which is realized by varying the power of the pump laser, and it follows the prediction provided from~\cite{laiho2023unfolding}. Our results show that a strong photon-pair correlation is necessary in order to achieve a high HOM dip visibility. Finally, we replace the WGMR with a new one made of the same material but with different dimensions and demonstrate that it is possible to match the temporal mode of the generated photons produced from two WGMRs.

\section{Results}

The experimental setup is schematically shown in Fig.~\ref{fig:experimental setup}. The WGMR we used in this experiment is the same as reported in Ref.~\cite{huang2024polarization}. It is coarsely temperature stabilized at \qty{90}{\degreeCelsius} by using a Peltier element and a temperature controller. We utilize a \qty{532}{\nm} continuous-wave laser as a pump laser and couple into the WGMR from the CW and CCW directions. The non-polarizing beamsplitters located on the left side of the WGMR are used to detect the reflected whispering gallery mode spectrum. One of this detected signals is also used as the input signal of Pound–Drever–Hall technique~\cite{black2001introduction} in order to generate an error signal for further stabilizing the WGMR~\cite{shafiee2020nonlinear}. The x-cut LiNbO\(_{3}\) prism is used as a selective coupler which brings pump light into the resonator while forbids the signal and idler to be coupled out.~\cite{sedlmeir2017polarization}. The diamond prism is used to couple the generated photons out of the WGMR. Then, the signals from both propagating directions are each guided to impinge on the input ports of a non-polarizing beamsplitter such that the HOM interference can happen. After that the light from the beamsplitter output ports are coupled into two detectors. The idler beams are coupled to other two detectors separately and used as heralds. A four-fold coincidence is registered when the four detectors have clicked within given time windows.

\begin{figure}[htbp]
    \includegraphics[width = \columnwidth]{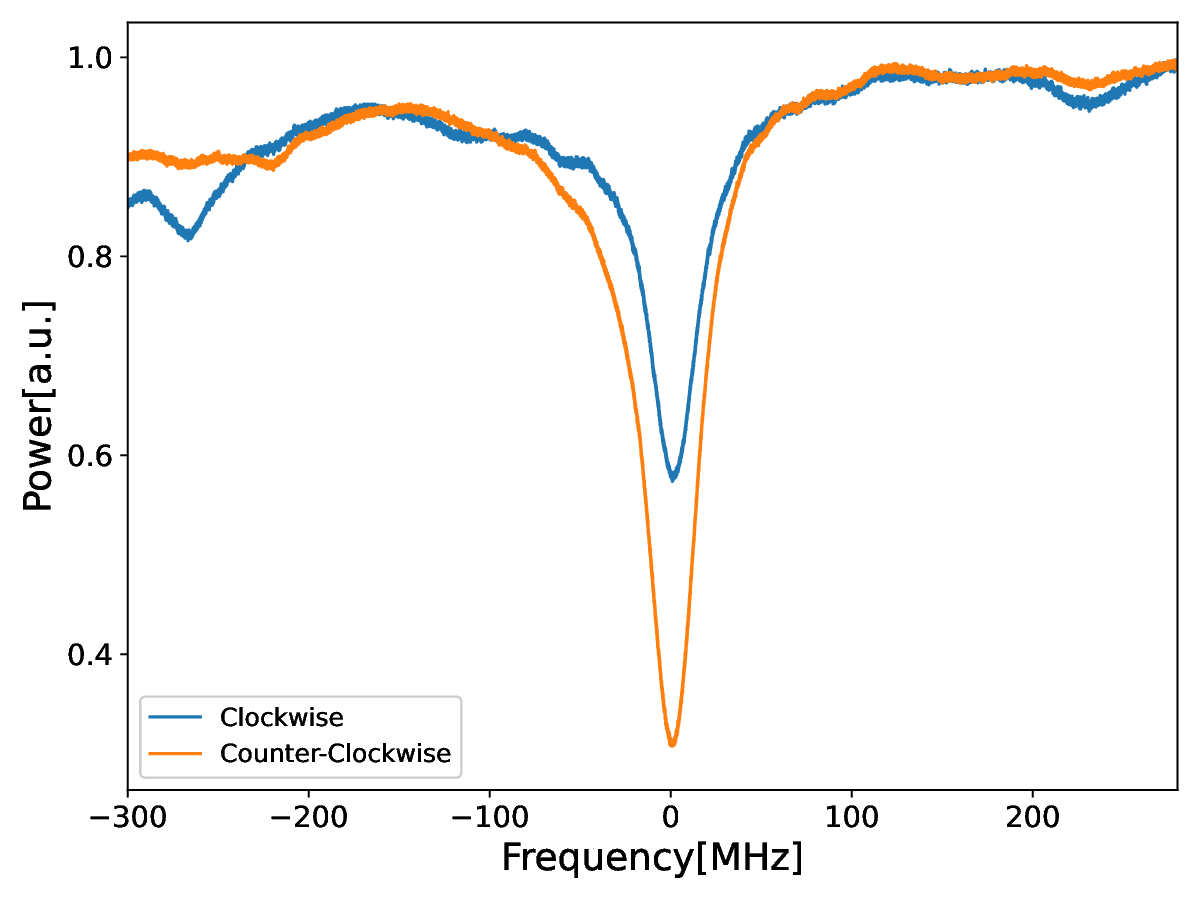}
    \caption{\label{fig:pump mode}Reflected pump spectra for the CW and CCW beams measured from D1 and D2, respectively. We use a wavemeter to measure the frequency sweep range of the pump laser and map it to the horizontal axis. The coupling efficiency is \(25\%\) for the CW and \(50\%\) for the CCW direction.}
\end{figure}

To check whether the pump beams traveling in both directions are coupled to the same whispering gallery mode, we measure the reflected spectra from both directions. If the two pump beams are coupled to the same whispering gallery mode, the reflected spectra are the same. We show the results in Fig.~\ref{fig:pump mode}. The reflected spectra have the same bandwidth of \qty{38}{MHz}, resonate at the same frequency, and have similar form, which confirms that the two beams occupy the same whispering gallery mode. The slight differences in the mode profiles and coupling efficiencies are due to imperfections in the optical components we used in our experiment. For example, the LiNbO\(_{3}\) prisms we use here are not isosceles, so when we couple the light from both sides, we get different astigmatism, which results in different spectra and coupling efficiencies.

\begin{figure}[htbp]
    \includegraphics[width = \columnwidth]{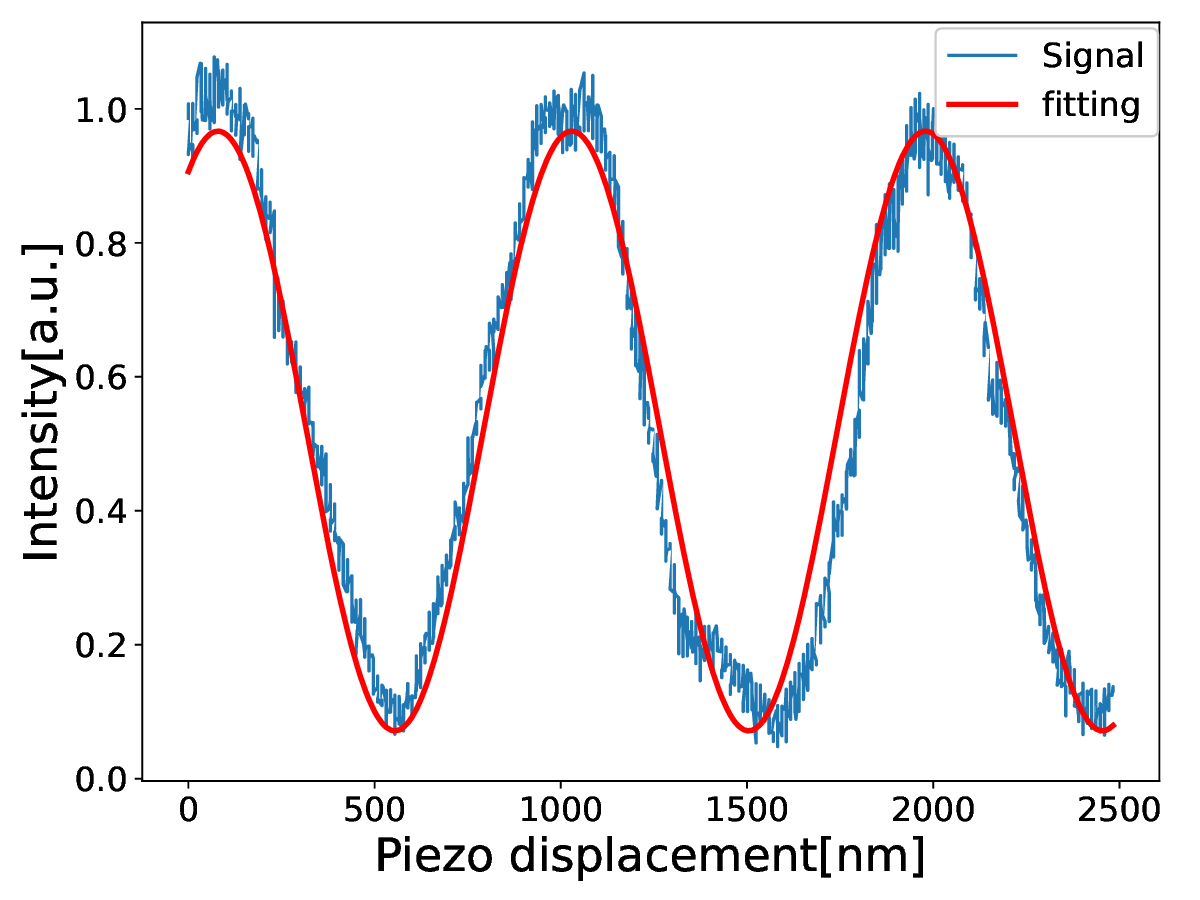}
    \caption{\label{fig:classical interference} Normalized interference pattern on the signal beams' paths in OPO regime. The blue line is the experimental result and the red line is the fitting result of a sinusoidal wave. The visibility is defined as \((I_{\mathrm{max}} - I_{\mathrm{min}})/(I_{\mathrm{max}} + I_{\mathrm{min}})\), which is \(86\%\) in this case.}
\end{figure}

The spatial mode overlap and the polarization likeliness are optimized based on classical interference. To this end, we increase the pump power in order to drive our WGMR in the optical parametric oscillator~(OPO) regime, where parametric photons are transformed into coherent light and the interference pattern on the signal beams' paths can be measured after combining the light from both directions on a non-polarizing beamsplitter. Note that the phase locking between the CW and CCW signal modes responsible for the stationary interference fringe in Fig.~\ref{fig:classical interference} is not a trivial phenomenon. It is not expected for perfectly independent CW and CCW OPOs, and is likely to originate from the signals and idlers back-scattering. By maximizing the visibility of the classical interference, the spatial- and polarization-mode indistinguishability are optimized as shown in Fig.~\ref{fig:classical interference}. The highest visibility we achieved in this experiment is \(86\%\). The limiting factor comes mainly from the spatial mode distortion caused by the optical components we used. The visibility can be easily improved to almost unity if we first couple the signals into fibers and then let them to interfere.

\begin{figure}[htbp]
    \includegraphics[width = \columnwidth]{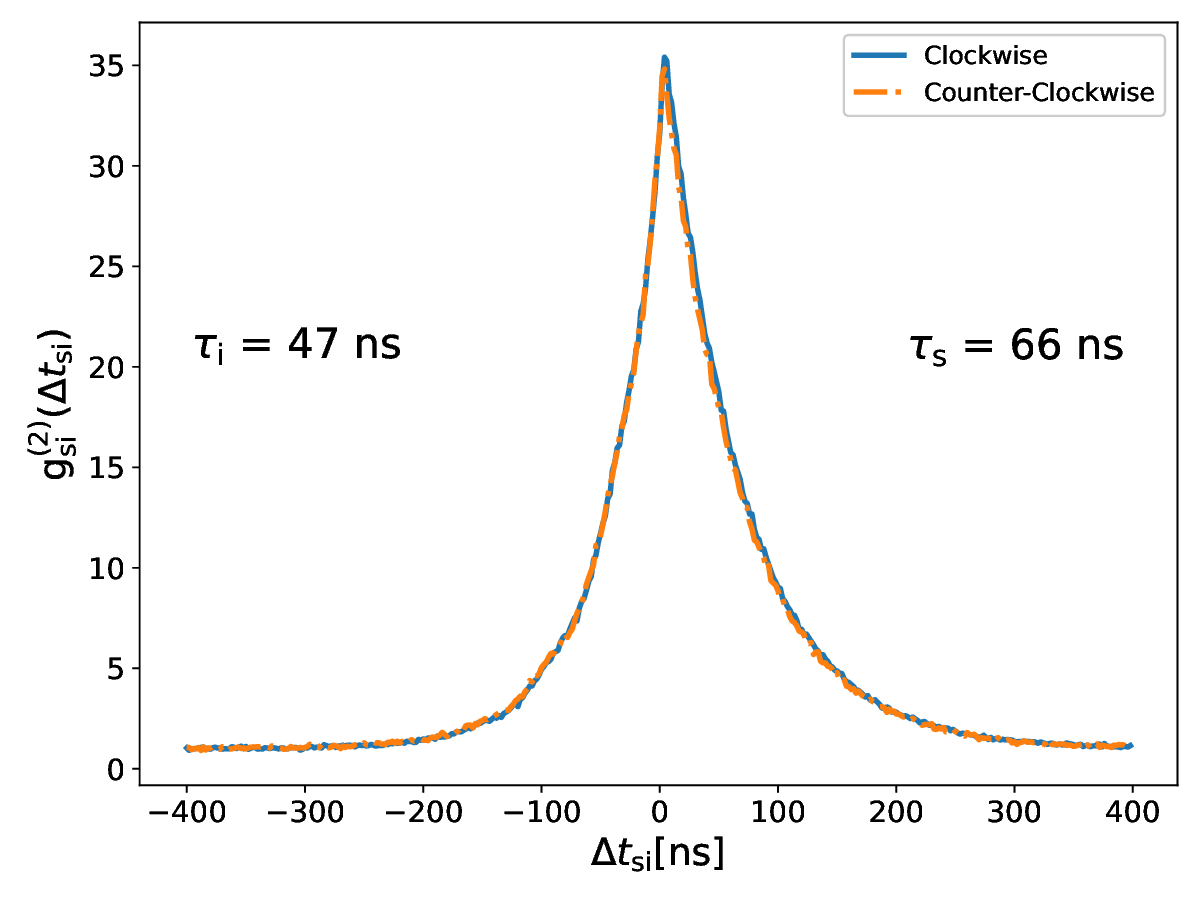}
    \caption{\label{fig:g2}Normalized second-order correlation functions taking the value \(g^{(2)}_{\mathrm{si}}(0)=35\). The decay constants are \qty{47}{ns} for the lead part, and \qty{66}{ns} for the tail part. The difference is due to the difference of the linewidth between the signal and idler~\cite{huang2024polarization}.}
\end{figure}

The similarity of the temporal modes between the counterpropagating heralded photons 
are confirmed by measuring the cross-correlation functions in both directions. The results are shown in Fig.~\ref{fig:g2}. We use double exponential decay function to fit the results and use the fitted results to calculate the similarity between two cross-correlation functions. The similarity \(S\) is defined as

\begin{equation}
    \label{eq:similarity}
    S=\frac{\int_{t_{\mathrm{c}}} f(x)g(x) \,dx}{\sqrt{\int_{t_{\mathrm{c}}} f(x)^2 \,dx \times \int_{t_{\mathrm{c}}} g(x)^2 \,dx}},
\end{equation}
where \(t_{\mathrm{c}}\) is the coherence time that is defined as the relative time at which \(g^{(2)}_{\mathrm{si}}(\Delta t_{\mathrm{si}})\) decayed by \(1 / \mathrm{e}^2\).
For two identical functions, it is easy to see that the value \(S\) is equal to 1. In our experiment, the value \(S\) is \(99.99\%\), showing that the temporal mode of the parametric photons between both propagating directions are well matched.

\begin{figure}[htbp]
    \includegraphics[width = \columnwidth]{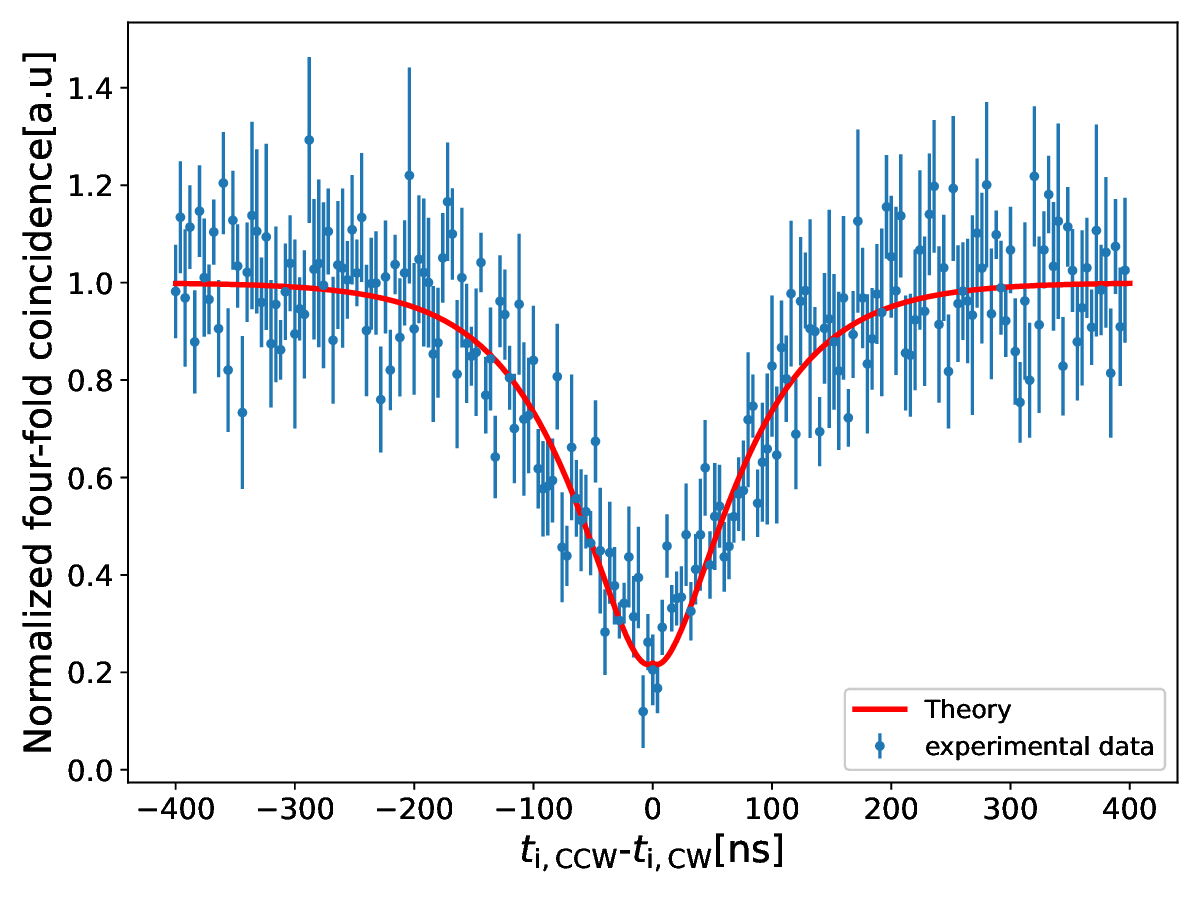}
    \caption{\label{fig:hom}Normalized HOM interference measured in four hours. The values of \(g^{(2)}_{\mathrm{si}}(0)\) are 35 both in the CW and CCW directions. The red line presents the theoretical predication including the parameters extracted from Fig~\ref{fig:g2}. The errors are calculated assuming that the main uncertainties come from the photon counting statistics, and that the statistics are Poissonian.}
\end{figure}

Fig.~\ref{fig:hom} shows the investigated HOM interference and the theoretical prediction which is calculated according to Ref.~\cite{laiho2023unfolding} with the photon spectrum taken from Ref.~\cite{chuu2011ultrabright}. We assume that the single round-trip power loss in the crystal is negligible and the splitting ratio of the non-polarizing beamsplitters is 0.5. The efficiencies on each beam path required for the model in the Ref.~\cite{laiho2023unfolding} are estimated based on experimental data. We record the four-fold coincidence dip with only \qty{50}{nW} in-coupled pump power in each propagating directions. The four-fold coincidences are extracted with following steps. First, we record all the events in which the two idler detectors click within time \(\Delta t\). In order to observe the whole shape of the HOM interference dip, \(\Delta t \) should be larger than the photon coherence time. In our case, we choose \(\Delta t\) as \(\pm \qty{400}{ns}\). We then look for the clicks of two signal detectors in a given coincidence window related to the coherence time of the signal and idler, each click corresponding to one of the idler detectors. From Fig.~\ref{fig:g2}, we know the coincidence window is not symmetric but ranges from \(-2\tau_{\mathrm{i}}\) to \(2\tau_{\mathrm{s}}\). This measurement technique has advantages over the traditional HOM measurement when \(\Delta t\) is fixed by the beam splitter position, which needs to be varied in order to obtain an equivalent of Fig.~\ref{fig:hom}. In our case, moving the beam splitter is not necessary and in fact would not be efficient, considering the long coherence length of the WGMR SPDC photons. Instead, all data points in Fig.~\ref{fig:hom} are recorded in a single measurement with fixed experimental setup, and then postselectively sorted into a time series. The visibility of the HOM interference is defined as

\begin{equation}
    V=\frac{C(\Delta t \rightarrow \infty) - C(\Delta t = 0)}{C(\Delta t \rightarrow \infty)},
\end{equation}

The visibility \(V\) gained from the results in Fig.~\ref{fig:hom} is \(74 \pm 5\%\). Because of non-perfect spatial overlap (only 86\% spatial overlap in our case) and the low \(g^{(2)}_{\mathrm{si}}(0)\) value, the visibility is limited. The spatial overlap can be improved by realizing the HOM interference with fiber-integrated optics, while the \(g^{(2)}_{\mathrm{si}}(0)\) value can be increased by reducing the pump power or decreasing the distance between the coupling prisms and the resonator.~\cite{fortsch2013versatile}

\begin{figure}[htbp]
    \includegraphics[width = \columnwidth]{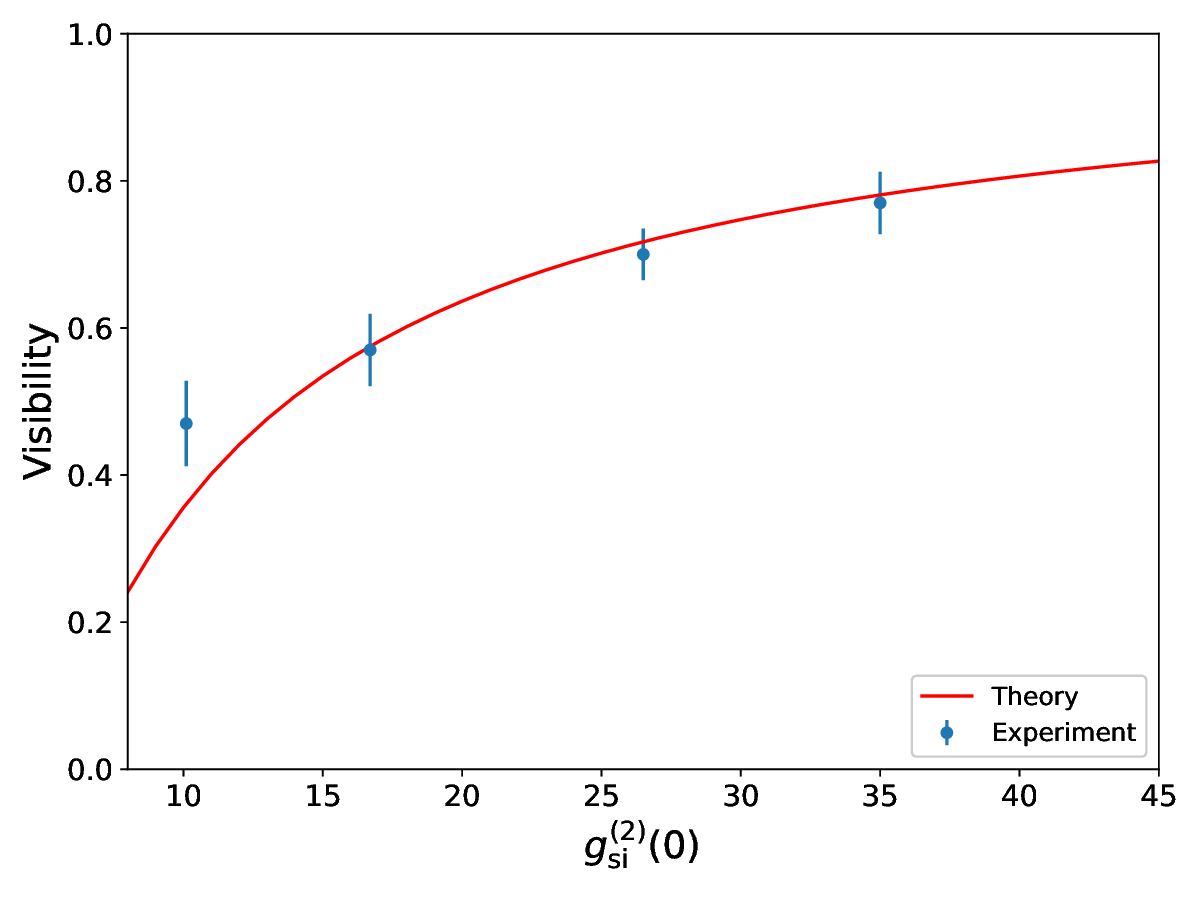}
    \caption{\label{fig:g2_vs_vis}Visibility of the HOM interference in terms of the peak value of the signal and idler cross-correlation function. The errors are calculated assuming that the main uncertainties come from the photon counting statistics, and that the statistics are Poissonian.}
\end{figure}

In order to see the relation between \(g^{(2)}_{\mathrm{si}}(0)\) and the visibility of the HOM dip, we change the pump power to vary the value of \(g^{(2)}_{\mathrm{si}}(0)\). Fig.~\ref{fig:g2_vs_vis} shows the experimental results and theoretical prediction, that agree nicely. One interesting point is that when the \(g^{(2)}_{\mathrm{si}}(0) = 10\), the visibility of HOM interference is larger than the theoretical prediction. The reason for this is that the theoretical prediction is calculated assuming that the SPDC process is at low-gain regime. This assumption works well at low pump powers. However, when the pump power increases (the value of \(g^{(2)}_{\mathrm{si}}(0)\) decreases), we are leaving the low-gain regime. In this case, this assumption does not hold and therefore the theoretical prediction does not work well for this data point.

\begin{figure}[htbp]
    \includegraphics[width = \columnwidth]{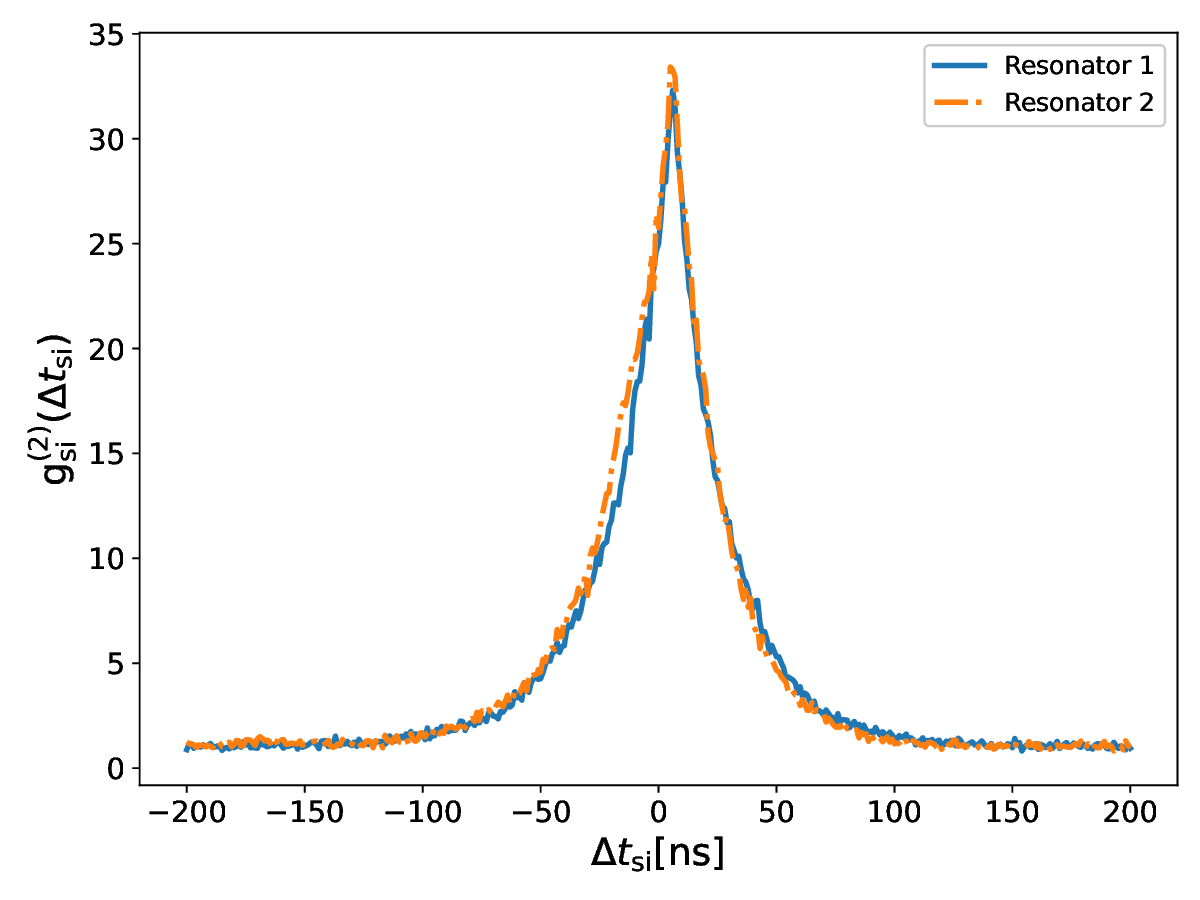}
    \caption{\label{fig:g2_two_resonators} Normalized second\orange{-}order correlation functions from two WGMRs. Resonator 1 is the one we used for all measurements here, and Resonator 2 is the new resonator.}
\end{figure}

Finally, we demonstrate that it is possible to match the temporal mode of heralded photons from different WGMR by fabricating another resonator of different size using the same material and comparing the temporal mode with the resonator used above. The new resonator has a radius \(R\approx \qty{0.75}{mm}\) with a rim radius \(r \approx \qty{0.14}{mm}\). We use the same \qty{532}{nm} laser as the pump laser and tune the wavelength of the signal to \qty{950}{nm}. In order to match the temporal mode, we vary the distance of the coupling prisms to change the bandwidths of SPDC photons~\cite{fortsch2013versatile}. The results of two temporal modes of heralded photons from two WGMRs are shown in Fig.~\ref{fig:g2_two_resonators}. The similarity S between these two temporal modes are \(99.97\%\), indicating that the temporal modes are well matched. Note that the frequency of generated photons has been shown to be continuously tunable by perturbing the evanescent field~\cite{schunk2015interfacing}, or applying electric field on the resonator~\cite{minet2022electro}. This paves the way for an efficient HOM interference from different WGMRs, which is necessary for large-scale quantum applications.

In conclusion, we have exploited the bi-directional pumping scheme to generate two pairs of indistinguishable photons in a WGMR and demonstrate the HOM interference between heralded states by counting four-fold coincidences. The highest visibility we achieve is \(74 \pm 5\%\). We experimentally show the relation between the \(g^{(2)}_{\mathrm{si}}(0)\) and the visibility of the HOM interference. The results are consistent with the theoretical prediction in low-gain regime. We also show that the temporal modes of the heralded photons from two WGMRs overlap very well, proving that it is possible to generate identical heralded photons from different WGMRs. Combining this with ultra low pump power requirement, we believe that WGMRs are promising platforms for realizing large scale quantum information applications.

\begin{acknowledgments}
This research was conducted within the scope of the project QuNET, funded by the German Federal Ministry of Education and Research (BMBF) in the context of the federal government's research framework in IT-security "Digital. Secure. Sovereign".
\end{acknowledgments}

\bibliography{paper_draft}

\end{document}